\pdfoutput=1 
\documentclass[aps,pra,twocolumn,groupedaddress,amsmath]{revtex4-1}

\usepackage{graphicx}
\usepackage{amssymb}
\usepackage{bm}
\usepackage{amsmath,amsfonts,latexsym}

\usepackage{braket} 

\usepackage{color}

\usepackage{comment}

\usepackage{bm} 

\newcommand*{\br}{\mathbf{r}}

\newcommand*{\cE}{{\cal E}}

\begin{document}

\title{Persistent Currents in Ferromagnetic Condensates}

\author{Austen Lamacraft}
\affiliation{TCM Group, Cavendish Laboratory, University of Cambridge, J. J. Thomson Ave., Cambridge CB3 0HE, UK}
\date{\today}
\email{al200@cam.ac.uk}

\date{\today}

\begin{abstract}

Persistent currents in Bose condensates with a scalar order parameter are stabilized by the topology of the order parameter manifold. In condensates with multicomponent order parameters it is topologically possible for supercurrents to `unwind' without leaving the manifold. We study the energetics of this process in the case of ferromagnetic condensates using a long wavelength energy functional that includes both the superfluid and spin stiffnesses. Exploiting analogies to an elastic rod and rigid body motion, we show that the current carrying state in a 1D ring geometry transitions between a spin helix in the energy minima and a soliton-like configuration at the maxima. The relevance to recent experiments in ultracold atoms is briefly discussed.

\end{abstract}

\maketitle

\section{Introduction} 
\label{sec:introduction}

Long-lived metastable currents -- often called \emph{persistent currents} -- are a hallmark of superfluidity \cite{Leggett:2000}. In Bose condensates, the existence of persistent currents is usually connected with the topology of the condensate wavefunction $\Psi(\br)$. Suppose that $\Psi(\br)$ varies only around the circumference of an annular container. If the vanishing of the superfluid density $\rho(\br)=|\Psi(\br)|^2$ is energetically unfavorable -- due to repulsive interparticle interactions or Fermi pressure in the case of a paired fermion superfluid -- only the phase of $\Psi(\br)$ may vary, winding an integer number of times around the circumference. The superfluid velocity is related to the gradient of the phase $\chi(\br)$ by
\begin{align}
  \mathbf{v} = \nabla\chi,
\end{align}
where the mass and $\hbar$ are set to unity. Winding of the phase is thus associated with current carrying states, metastable due to the energetic cost of vanishing density.

In this paper we are concerned with Bose condensates with internal degrees of freedom \cite{Stamper-Kurn:2013aa}, where the persistent currents can have a very different character. A particle may be in a superposition of different internal states, which we will generically refer to as `spin' states. To illustrate the physics involved, consider the simplest case of two internal states (`spin-1/2'). The order parameter is now a two component spinor. Consider the family of order parameter configurations~\cite{Leggett:2000}
\begin{align}
  \Psi(x) = \begin{pmatrix}\psi_\uparrow(x) \\\psi_\downarrow(x)\end{pmatrix}=\begin{pmatrix}
\cos(\theta/2)e^{2\pi ix/L}\\
\sin(\theta/2)
\end{pmatrix},\, 0\leq \theta\leq \pi
\end{align}
where $L$ is the circumference of the annulus. As $\theta$ increases from $0$ to $\pi$, we interpolate smoothly from a configuration where the phase winds by $2\pi$ to one with no winding. The density $\rho(x) = \Psi^\dagger\Psi$ is constant, so there is no energy penalty associated with compressing the condensate. Furthermore, the kinetic energy descreases monotonically.

This does not imply, however, that there are no metastable states in spinor condensates. If the `magnetization'
\begin{equation}
  M \equiv \int_0^L dx \left[\psi^\dagger_\uparrow\psi_\uparrow-\psi^\dagger_\downarrow\psi_\downarrow\right]
\end{equation}
is conserved, states of different $\theta$ in the above example are not dynamically accessible, as $M\propto \cos\theta$. Then it is clear that the fully polarized condensate behaves as its scalar counterpart. Only a partially polarized condensate can take advantage of the spin degrees of freedom to `unwind' the superfluid flow without changing the density. Our goal is to analyze the energy barrier present in this case.

 In a recent experiment \cite{Beattie:2013aa}, persistent currents were observed in a two-component Bose condensate confined to a toroidal trap. We will discuss this experiment in light of our results in Section \ref{sec:conclude}. Refs.~\cite{Abad:2014,Wu:2015} provide earlier treatments complimentary to ours.

\subsection{Energy Functional}

We will follow the approach introduced in Ref.~\onlinecite{lamacraft2008}, in which the mean field (or Gross-Pitaevskii) approximation is specialized to the `incompressible limit' where the interaction energy is \emph{much larger} than the kinetic energy of the superfluid flow. This is equivalent to the condensate wavefunction varying on a scale far in excess of the healing length.

We briefly recapitulate the key ideas for spin-1/2. The interaction energy is minimized by configurations of constant density $\rho$.
The variation of the energy between configurations of constant density arises \emph{solely} from the kinetic energy. If the transverse dimensions of a ring of circumference $L$ are much smaller than the healing length, this is
\begin{equation}
  \cE = \frac{1}{2}\int_0^L dx\, \frac{d\Psi^\dagger}{dx}\frac{d\Psi}{dx}.
\end{equation}
If we parameterize the spinor as
\begin{equation}
\Psi=\sqrt{\rho}e^{i\chi}\begin{pmatrix}
\cos(\theta/2)\\
\sin(\theta/2)e^{i\phi}\\
\end{pmatrix},
\end{equation}
then the angles $\theta$ and $\phi$ give the polar coordinates of a unit vector $\mathbf{n}$ giving the magnetization axis
\begin{equation}
  \rho\mathbf{n} = \Psi^\dagger\boldsymbol{\sigma}\Psi = \rho(\sin\theta\cos\phi,\sin\theta\sin\phi,\cos\theta),
\end{equation}
while the superfluid velocity is
\begin{align}
  v &= -\frac{i}{2\rho}\left[\Psi^\dagger \frac{d\Psi}{dx} - \frac{d\Psi^\dagger}{dx}\Psi\right]\\
  &=\chi' + s\phi'(1-\cos\theta)
  \label{eq:velocity}
\end{align}
where $s=1/2$. The energy can then be written as a functional of $\mathbf{n}(x)$ and $\chi(x)$
\begin{equation}
\cE[\mathbf{n},\chi] = \rho\int_0^L dx\left[\frac{v^2}{2} + \frac{s}{4}\left(\frac{d\mathbf{n}}{dx}\right)^2\right]
\label{eq:energy_fun}
\end{equation}
Since the angles $\chi$, $\theta$, and $\phi$ are all periodic on the ring, the supercurrent has circulation
\begin{equation}
  \int_0^L v \,dx= 2\pi p +s\Omega[\mathbf{n}],
\label{eq:circulation}
\end{equation}
where $n$ is an integer and $\Omega[\mathbf{n}]$ is the solid angle enclosed by the path traced out by $\mathbf{n}(x)$ on the unit sphere
\begin{equation}
  \Omega[\mathbf{n}] = \oint (1-\cos\theta)d\phi
  \label{eq:Omega}
\end{equation}
The generalization to ferromagnetic states of arbitrary spin $s$ was considered in Ref.~\onlinecite{lamacraft2008}. Allowing for rotationally invariant interactions between spin-$s$ particles, the mean field phase diagram always includes a region where the condensate wavefunction is a spin coherent state.


%
The form \eqref{eq:velocity} of the superfluid velocity represents one parameterization of the coherent states $\ket{\mathbf{n}}$: in general the velocity may be written in terms of the Berry potential $A(\mathbf{n})\equiv -i\bra{\mathbf{n}}(d\ket{\mathbf{n}}/dx)$ as
\begin{equation}
  v =\frac{d\chi}{dx} +A(\mathbf{n})
  \label{eq:velocity_general}
\end{equation}
The circulation is however gauge invariant and given by \eqref{eq:circulation}.

The analysis of the functional \eqref{eq:energy_fun} is the goal of this paper. As we have discussed, we are interested in minimizing the energy at fixed magentization $M$. This can be implemented with a Lagrange multiplier, giving the functional
\begin{equation}
  \cE_\mathbf{h}[\mathbf{n},\chi] = \cE[\mathbf{n},\chi] - \int_0^L \mathbf{h}\cdot \mathbf{n}\, dx.
  \label{eq:eh}
\end{equation}

\subsection{Conserved supercurrent}\label{sec:conserved}

A simple first observation is that $\chi$ is a cyclic coordinate in the language of Lagrangian mechanics: only its derivative appears in \eqref{eq:energy_fun}. As a result the Euler--Lagrange equation for $\chi$ implies
\begin{equation}
  v(x) = \frac{d\chi}{dx} +A(\mathbf{n}) = \mathrm{const.},
\end{equation}
so that stationary points of the energy have spatially constant superflow. The circulation is thus simply proportional to $v$ and \eqref{eq:circulation} gives
\begin{equation}
  v = \frac{2\pi p}{L} + \frac{s\Omega[\mathbf{n}]}{L}.
\label{eq:current_omega}
\end{equation}
Thus we may extremize only the second term in \eqref{eq:energy_fun} subject to fixed $\Omega[\mathbf{n}]$.

The stationary configurations of the energy with fixed $v$, or total momentum $P=\rho v L$, move at constant velocity. To see this, consider the action describing the dynamics of the condensate
\begin{equation}
  S = \int dt \left[\int_0^L i\Psi^\dagger\partial_t\Psi\, dx - \cE[\mathbf{n},\chi]\right].
\end{equation}
For a configuration $\Psi(x,t)=\Psi(x-ut)$ moving at constant velocity $u$, the action is $S=-\int \cE_{u} dt $, where
\begin{equation}
  \cE_u = \cE - u P.
  \label{eq:Eu}
\end{equation}
Thus we can interpret $u$ as a Lagrange multiplier fixing $P$, so that
\begin{equation}
u = \frac{\partial \cE}{\partial P},
\label{eq:disp_deriv}
\end{equation}
as one might have expected.

On account of the conserved supercurrent, the dispersion relation we seek has the general form
\begin{equation}
\cE_p(\Omega,M) =  \frac{\rho}{2L} (2\pi p + s\Omega)^2 + \cE_\sigma(\Omega, M)
\label{eq:dispersion_total}
\end{equation}
where
\begin{equation}
  \cE_\sigma(\Omega, M) = \min_{\substack{\mathbf{n}(x)\\ \Omega[\mathbf{n}]=\Omega \\ M[\mathbf{n}]=M}} \frac{\rho s}{4}\int_0^L\left(\frac{d\mathbf{n}}{dx}\right)^2 dx.
  \label{eq:const_min}
\end{equation}
As shown in Ref.~\onlinecite{Tjon:1977}, the function $\cE_\sigma(\Omega, M)$ gives the dispersion relation of constant velocity solutions of the Landau--Lifshitz equation
\begin{equation}
  	\partial_{t}\mathbf{n}=\frac{\rho s}{2}\mathbf{n}\times\partial_{x}^{2}\mathbf{n}.
\end{equation}
\begin{figure}
  \includegraphics[width = \columnwidth]{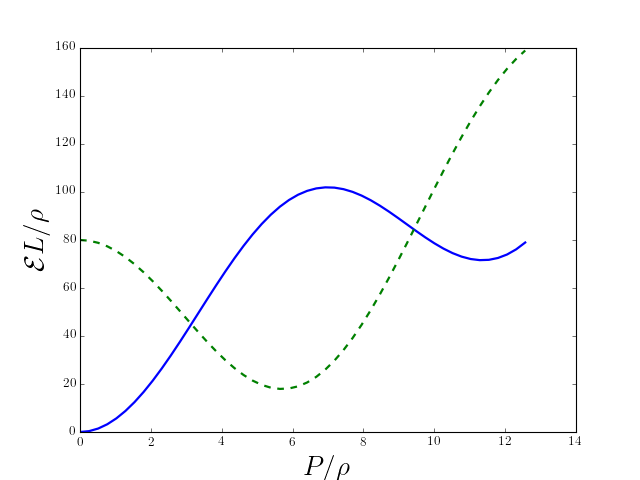}
  \caption{Two branches of the dispersion for $s=1$, with the solid and dashed lines corresponding to $p=0,1$ in \eqref{eq:dispersion_total}. We use the soliton dispersion \eqref{eq:tw_dispersion} at $\Delta=0.1$ for simplicity, see Section \ref{sec:soliton}.}
  \label{fig:2branches}
\end{figure}
The extra term in \eqref{eq:dispersion_total} represents the contribution of the superfluid velocity, absent in a normal magnet.  Since $\cE_\sigma(\Omega,M)$ has period $4\pi$ in $\Omega$, there are $2s$ separate `branches' of the dispersion indexed by $p=0,\ldots, 2s-1$, see Figure \ref{fig:2branches}. These branches have the following topological meaning. For $s=1/2$ constant density states satisfy
\begin{equation}
  |\psi_\uparrow|^2+|\psi_\downarrow|^2=1
\end{equation}
and correspond to points on the three dimensional sphere $S^3$. On $S^3$ there are no noncontractible loops, so that any periodic $\Psi(x)$ may be smoothly deformed to any other.

For $s>1/2$, we can parameterize the coherent state in terms of the $s=1/2$ spinor that has the same magnetization axis $\mathbf{n}$, but we have to be careful with the overall phase. When rotated through an angle $\theta$ about the $\mathbf{n}$ axis, the coherent state acquires a phase factor $e^{i\theta s}$. Thus we have to identify spin-1/2 states that differ by $e^{i\pi/s}$.
\begin{equation}
  \psi_{\uparrow,\downarrow}\sim e^{i\pi/s}\psi_{\uparrow,\downarrow}
\end{equation}
This quotient of $S^3$ is known as the \emph{Lens space} $L(p;1)$ and has fundamental group $\mathbb{Z}/(2s)$ \cite{Wikipedia:2016aa}\footnote{We thank Nick Manton for helping to clarify this point}. The $2s$ elements correspond to the $2s$ branches of the dispersion. The same physics was discussed in Ref.~\onlinecite{Haldane:1986aa} in the context of the Heisenberg ferromagnetic chain.

\subsection{Outline}

To close this introductory section, we outline the remainder of the paper. In the next section we introduce two precise mathematical analogies that give different interpretations of the functional \eqref{eq:energy_fun}, and provide a useful basis for further analysis. In Section \ref{sec:disp} we find the analytic form of the mininum energy solutions. The parameters of the solutions are fixed by the periodic boundary conditions. In Section \ref{sec:numerical} we give an alternative numerical formulation that is often more convenient in practice than the analytical solution. Finally, in Section \ref{sec:conclude} we discuss the relation of these results to recent experiments and other theoretical works.

\section{Two Analogies}\label{sec:2_analogies}

\subsection{Elastic Analogy}

We will now show that the energy functional \eqref{eq:energy_fun} also describes the elastic energy of a flexible rod. It is useful to first write the functional in a different way, motivated by the spin-1 case rather than the spin-1/2 context in which we introduced it. To this end, write the three component order parameter of a spin-1 condensate in term of two real vectors as
\begin{equation}
  \Psi = \sqrt{\frac{\rho}{2}}\left(\mathbf{a} + i\mathbf{b}\right).
\end{equation}
In a cartesian basis, the spin-1 operators have the explicit form $(S_j)_{kl}=-i\epsilon_{jkl}$, so that
\begin{equation}
  \Psi^\dagger\mathbf{S}\Psi = \rho \mathbf{a}\times\mathbf{b}.
\end{equation}
As a result, a fully polarized ferromagnetic condensate is described by orthogonal unit vectors $\mathbf{a}$, $\mathbf{b}$, with magnetization in the direction $\mathbf{n}=\mathbf{a}\times\mathbf{b}$. The orthonormal triad  $\{\mathbf{a}, \mathbf{b}, \mathbf{n}\}$ defines an element of the group of proper rotations $SO(3)$ \cite{Ho:1998aa,Ohmi:1998aa}. The two branches of the dispersion in this case correspond to the well known fact that $\pi_1(SO(3))=\mathbb{Z}/2$


In this language, the superfluid velocity has an interesting interpretation:
\begin{align}
  v &= -\frac{i}{2\rho}\left[\Psi^\dagger \frac{d\Psi}{dx} - \frac{d\Psi^\dagger}{dx}\Psi\right]\\
  &=\frac{1}{2}\left[\mathbf{a}\cdot \frac{d\mathbf{b}}{dx} - \frac{d\mathbf{a}}{dx}\cdot \mathbf{b}\right].
\end{align}
We see that $v$ describes the twisting around the $\mathbf{n}$ axis.



By using the relations
\begin{align}
    \mathbf{n}' = \mathbf{a}'\times\mathbf{b} + \mathbf{a}\times \mathbf{b}'\\
    \mathbf{a}'\cdot\mathbf{b}+ \mathbf{a}\cdot\mathbf{b}'=0,
\end{align}
we can obtain
\begin{align}
  \cE &= \frac{1}{2}\int_0^L  dx\,\frac{d\Psi^\dagger}{dx}\frac{d\Psi}{dx}\\
  &= \frac{\rho}{4}\int_0^L \left[\mathbf{a}'^2+\mathbf{b}'^2\right]\\
  &= \rho\int_0^L dx\left[\frac{v^2}{2} + \frac{1}{4}\left(\frac{d\mathbf{n}}{dx}\right)^2\right],
\end{align}
which coincides with \eqref{eq:energy_fun} for $s=1$.

\begin{figure}
  \includegraphics[width = 0.8\columnwidth]{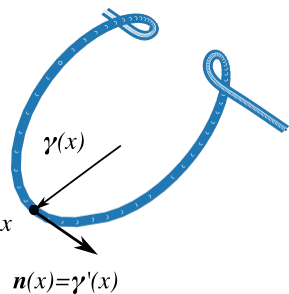}
  \caption{An elastic rod described by a space curve $\boldsymbol{\gamma}(x)$ as a function of arc length $x$, with unit tangent vector $\mathbf{n}\equiv\boldsymbol{\gamma}'(x)$. For this curve, the corresponding $\mathbf{n}(x)$ is shown in Figure \ref{fig:sphere}.}
  \label{fig:rod}
\end{figure}

Now we will see how the same formulation in terms of an orthonormal triad describes an elastic rod, by which we mean an elastic body with cross section small compared to its length. The rod's configuration is described by a curve $\boldsymbol{\gamma}(x)$ giving the position of a point on the centre line as a function of arc length $x$ (see Figure \ref{fig:rod}). $\mathbf{n}\equiv\boldsymbol{\gamma}'(x)$ is the unit tangent vector to the curve. One contribution to the energy of a configuration depends on the \emph{curvature} of the rod, which is
\begin{equation}
  \kappa = \left\lvert\frac{d\mathbf{n}}{dx}\right\rvert.
\end{equation}
The other contribution depends on the twisting of the rod. To quantify this, we need to set up an orthonormal triad $\{\mathbf{n}(x),\mathbf{a}(x),\mathbf{b}(x)\}$ at each point on the curve. As we move along the curve, this triad undergoes a rotation, which can be described by an `angular velocity' $\boldsymbol{\omega}$ (having units of inverse length)
\begin{align}
    \mathbf{n}'&=\boldsymbol{\omega}\times\mathbf{n}\\
    \mathbf{a}'&=\boldsymbol{\omega}\times\mathbf{a}\\
    \mathbf{b}'&=\boldsymbol{\omega}\times\mathbf{b}.
    \label{eq:omega_def}
\end{align}
If we write
\begin{equation}
  \boldsymbol{\omega}= \omega_n \mathbf{n}+\omega_a \mathbf{a}+\omega_b \mathbf{b},
\end{equation}
then the simplest approximation to the elastic energy of the rod is
\begin{equation}
  \cE_\text{el} = \frac{1}{2}\int dx\left[\alpha \omega_a^2+\beta\omega_b^2+ \gamma\omega_n^2\right].
  \label{eq:elastic_energy}
\end{equation}
The first two terms represent the energy of bending, and the third the energy of twisting. For a symmetric cross section $\alpha=\beta$, and the definition \eqref{eq:omega_def} can be used to write
\begin{align}
  \left(\frac{d\mathbf{n}}{dx}\right)^2 = \omega_a^2+\omega_b^2\\
  \mathbf{a}'\cdot\mathbf{b}=-\mathbf{a}\cdot\mathbf{b}'=\omega_n
\end{align}
The elastic energy \eqref{eq:elastic_energy} then coincides with the energy functional \eqref{eq:energy_fun} of the ferromagnetic condensate with the identifications $\alpha=\rho s/2$, $\gamma = \rho s^2$. The bending stiffness $\alpha$ corresponds to the spin stiffness, while the torsional stiffness $\gamma$ corresponds to the superfluid stiffness.

There is a complete correspondence between the elastic and superfluid problems. For example, the analog of constant supercurrent is constant twist rate $\omega_n=\text{const.}$, which arises from the vanishing torque along the axis on any element of the rod.

The end-to-end displacement of the rod
\begin{equation}
  \boldsymbol{\gamma}(L)-\boldsymbol{\gamma}(0) = \int \mathbf{n}(x)\,dx
\end{equation}
corresponds to the total magnetization, while a force applied to the endpoints corresponds to a magnetic field, appearing in the energy $\cE_\mathbf{h}$ in \eqref{eq:eh}. Thus a straight rod with $|\boldsymbol{\gamma}(L)-\boldsymbol{\gamma}(0)|=L$ corresponds to a fully polarized condensate. Evidently such a rod may be twisted indefinitely about its axis (until the approximations of linear elasticity break down). This is the case of a fully polarized superfluid: an arbitrary number of phase twists are allowed in the constant density approximation.

If the ends of the rod are allowed to approach each other the rod bends, increasing the bending energy, but will tend to untwist, reducing the torsional energy. This corresponds to a partially polarized condensate, and is the mechanism by which the supercurrent is reduced. Note that a real elastic rod can touch itself when buckled, a feature that does not appear in the superfluid problem.

To quantify the untwisting associated with bending of the rod imagine fixing a twist $\psi$ along a straight rod. If the ends of the rod are clamped in the same relative orientation so that the tangents match $\mathbf{n}(0)=\mathbf{n}(L)$, then the relation \eqref{eq:current_omega} takes the form (for $s=1$)
\begin{equation}
  \int \omega_n\, dx = \psi - \Omega[\mathbf{n}].
  \label{eq:twist_integral}
\end{equation}
(note $v\to -\omega_n$) As the solid angle enclosed by $\mathbf{n}(x)$ increases, the integrated twist decreases.

We summarize the superfluid-elastic analogy in Table \ref{table:elastic}. The theory of elastic rods is extensive: we refer to the recent book \cite{Audoly:2010} for further details and references. The field begins with Kirchhoff's paper of 1859, which introduced the second analogy we will use \cite{Kirchhoff:1859aa}.

\begin{widetext}
\begin{table*}
\begin{tabular}{||c|c|c||}
  \hline
  Ferromagnetic Condensate & Elastic Rod & Lagrange Top\\
  \hline
  $\mathbf{n}$ & $\boldsymbol{\gamma}'(x)$, tangent to rod & Symmetry axis of top (3 axis) \\
  $|\mathbf{n}'|$ & $\kappa$, curvature & $\sqrt{\omega_1^2+\omega_2^2}$ \\
  $v$, supercurrent & $\omega_n$, twist rate & $\omega_3$ \\
  Superfluid stiffness (coefficient of $v^2$ in energy) & Torsional stiffness &  $I_3$, moment of inertial about 3 axis \\
  Spin stiffness (coefficient of $\mathbf{n}'^2$ in energy) & Bending stiffness & $I_1=I_2$ \\
  $\mathbf{M}$, magnetization  & $\boldsymbol{\gamma}(L)-\boldsymbol{\gamma}(0)$, end-to-end displacement & N/A \\
  $\mathbf{h}$, magnetic field & Force applied at endpoints & Gravity acting on 3 axis\\
  \hline
\end{tabular}
\caption{The elastic and kinematic analogies}
\label{table:elastic}
\end{table*}
\end{widetext}

\subsection{Kinematic Analogy}

Kirchhoff discovered a precise mapping between the equilibrium configurations of an elastic rod and the motion of a rigid body. With the groundwork of the previous section, it is not too hard to understand where this \emph{kinematic analogy} comes from. Arc length along the rod is interpreted as time, and we have already seen that the change of the orthonormal frame describing the configuration of the rod defines an angular velocity. The elastic energy \eqref{eq:elastic_energy} is now interpreted as an \emph{action} given by the time integral of the rigid body kinetic energy. The stiffnesses $\alpha$, $\beta$ and $\gamma$ are interpreted as moments of inertia $I_{1,2,3}$. The case $\alpha=\beta$ of interest to us corresponds to the \emph{symmetric top} with symmetry axis $\mathbf{n}$. The kinematic analogy is summarized in Table \ref{table:elastic}.

A Lagrangian that includes a linear potential acting on the symmetry axis corresponds to the energy $\cE_\mathbf{h}$ of \eqref{eq:eh}. Recalling that the magnetic field $\mathbf{h}$ plays the role of a Lagrange multiplier in our formulation, we see that the problem of ferromagnetic superfluids with fixed magnetization is in correspondence with the heavy symmetric top, often called the \emph{Lagrange top}. This system is treated in many textbooks, see e.g. Ref. \onlinecite{Jose:1998}. In terms of the usual Euler angles, and taking $\mathbf{h}$ to lie in the $z$-direction, the energy $\cE_\mathbf{h}$ is
%
%
\begin{multline}
  \cE_\mathbf{h} = \int_0^L dx\left[ \frac{\alpha}{2}\left(\theta'^{2}+ \phi'^{2}\sin^{2}\theta\right)+\frac{\gamma}{2}\left(\psi'+\phi'\cos\theta\right)^{2}\right.\\
  \left.-h\cos\theta\right]
\end{multline}
The Lagrange top is a completely integrable system on account of its three conserved quantities. The first two are the angular momentum about the symmetry axis, corresponding to the conserved supercurrent, and the angular momentum in the direction of the field. These are associated with the cyclic coordinates $\psi$ and $\phi$, and take the form
\begin{align}
  p_\psi &= \gamma\left(\psi'+\phi'\cos\theta\right)\\
  p_\phi &=\alpha\phi'\sin^{2}\theta+\cos\theta\, p_{\psi}.
  \label{eq:2p}
\end{align}
$p_\phi$ corresponds to the $z$-component of spin current. The third conservation law is associated with translation -- in the dynamical case this is the energy though here it has units of energy density
\begin{equation}
  \varepsilon \equiv \frac{\alpha}{2}\left(\theta'^{2}+ \phi'^{2}\sin^{2}\theta\right)+h\cos\theta.
\end{equation}
Using the first two conservation laws $\varepsilon$ may be expressed purely in terms of $z\equiv\cos\theta$ as
\begin{align}
  \varepsilon &= \frac{\alpha}{2}\frac{z'^{2}}{1-z^{2}} + \frac{\left(p_{\phi}-z p_{\psi}\right)^{2}}{2\alpha\left(1-z^{2}\right)}+hz.
  \label{eq:eps_cons}
\end{align}
In this way the dynamics is reduced to a single variable moving in an effective potential
\begin{equation}
  V_\text{eff}(z) = \frac{1}{2\alpha}\left(p_{\phi}-z p_{\psi}\right)^{2}+(hz-\varepsilon)(1-z^2).
  \label{eq:Veff}
\end{equation}
$V_\text{eff}(z)$ is a cubic polynomial. For $h>0$ motion will occur between the two smallest roots, which must lie in $[-1,1]$, while the third root is at $z\geq 1$.

The solution to \eqref{eq:eps_cons} can be found in terms of elliptic functions. We will use this solution in the next section to calculate the dispersion relation.

\section{Dispersion Relation}\label{sec:disp}

\subsection{Boundary Conditions}

A problem in dynamics is typically an initial value problem, where one is interested in the evolution that results from some initial conditions. In contrast, we are here concerned with a boundary value problem. What boundary conditions should we impose at $x=0,L$? It is important to remember that we are concerned with finding extremal configurations at fixed $P$, or equivalently extremizing the functional \eqref{eq:Eu}
\begin{equation}
  \cE_u = \cE - u P.
\end{equation}
We now show that this is equivalent to working at fixed $p_\psi$, ignoring the boundary conditions on $\psi$, and enforcing periodic boundary conditions only on the vector $\mathbf{n}$.
\begin{align}
  \theta(L)-\theta(0)&=0\nonumber\\
  \phi(L)-\phi(0) &= 2\pi n_\phi.
  \label{eq:pbc}
\end{align}
The reason is that any extremum of $\cE_u$ with periodic boundary conditions for $\psi$ can be gauge transformed (or canonically transformed in the language of classical mechanics) to an extremum of $\cE$ with twisted boundary conditions by the transformation
\begin{equation}
  \psi(x) \to \psi(x) + \frac{ux}{\gamma},
\end{equation}
after which
\begin{equation}
  \cE_u \to \cE - \frac{u^2}{2\gamma}.
\end{equation}
Recall that there are three conserved quantities in the solution discussed in the previous section. The two conditions \eqref{eq:pbc} therefore give a one parameter family of solutions (using $h$ to fix $M$). From each member of this family we can extract the momentum and energy
\begin{align}
  P/\rho &= 2\pi p + s\Omega\\
  \cE &= \varepsilon L - hM + \frac{ P^2}{2\rho L},
  \label{eq:EP_implicit}
\end{align}
where the solid angle is computed using \eqref{eq:Omega}. Note that $P\neq p_\psi L/\gamma$ on account of the twisted boundary conditions, but rather
\begin{equation}
  P = \frac{L}{\gamma}(p_\psi - u).
\end{equation}

\subsection{Spin helix}\label{sec:helix}

A particularly simple configuration corresponds to a spin helix, in which $\theta(x)=\theta_0$ while $\phi(x)$ increases by a multiple of $2\pi$. In the kinematic analogy, this corresponds to simple precession of the top. This occurs when the two smallest roots of the cubic potential \eqref{eq:Veff} coincide. In this case $\phi'(x)$ and $\psi'(x)$ are both constant and the energy is simply
\begin{equation}
  \cE =  \frac{2\pi^2n_\phi^2\alpha }{L} (1-m^2)+\frac{P^2}{2\rho L},
  \label{eq:helix_energy}
\end{equation}
where $\phi'=2\pi n_\phi/L$ and $\cos\theta_0 = M/L\equiv m$. Note that the solid angle and polarization are related by
\begin{equation}
\Omega_\text{c}=\Omega = 2\pi (1-\cos\theta_0)=2\pi (1-m),
\label{eq:Omega_helix}
\end{equation}
so this configuration represents a single point on the dispersion relation. If we fix the  $z$-component of the magnetization, a smaller momentum can be obtained from a `canted' helical configuration that winds around an axis tilted relative to the $z$-axis. Note that in this region the Lagrange multiplier $h=0$, which is why the orientation of the helix is free. The energy $\cE_\sigma(\Omega,M)$ is straightforward to evaluate
\begin{equation}
  \text{Helix: }\cE_\sigma(\Omega,M) = \frac{2\pi\alpha\Omega}{L}\left(1-\frac{\Omega}{4\pi}\right)
  \label{eq:helix_disp}
\end{equation}
As the solid angle increases to $\Omega_\text{c}$, this axis aligns with the $z$-axis. Thereafter, the only way to increase $\Omega$ at fixed $M$ is by forming a localized, soliton-like solution.

\subsection{Buckling of the helix}\label{sec:buckle}

For $\Omega>\Omega_\text{c}$ there is a qualitative change in the configuration: $\theta(x)$ begins to develop a nonuniform profile, corresponding to \emph{nutation} of the top. Assuming that the amplitude of this profile increases continuously from $0$ (as will be verified by the numerical calculations of Section \ref{sec:numerical}), we can treat the motion about $\theta_0 =\cos^{-1}(m)$ as harmonic and solve the quantization conditions analytically.

For a single period of $\theta(x)$ motion, we have the following conditions on the effective potential
\begin{align}
  \label{eq:Veff_eq}
  V_\text{eff}(m)&=V_\text{eff}'(m)=0\\
  V_\text{eff}''(m)&=\alpha \left(\frac{2\pi}{L}\right)^2.
\end{align}
For a single turn of the helix $n_\phi=1$ and
\begin{equation}
  p_\phi - m p_\psi = \frac{2\pi \alpha}{L}(1-m^2).
  \label{eq:p_rel}
\end{equation}
The four equations \eqref{eq:Veff_eq}-\eqref{eq:p_rel} determine the four quantities $p_\psi$, $p_\phi$, $\varepsilon$, and $h$. There are two sets of solutions. The first corresponds to the spin helix found before:
\begin{subequations}
\label{eq:sol_one}
\begin{align}
p_\psi L = 2\pi\alpha m,\qquad p_\phi L=2\pi\alpha,\\ \varepsilon L^2 =2\alpha\pi^2(1-m^2),\qquad hL^2=0,
\end{align}
\end{subequations}
while the second is
\begin{subequations}
\label{eq:sol_two}
\begin{align}
p_\psi L = 6\pi\alpha m,\qquad p_\phi L=2\pi\alpha(1+2m^2),\\ \varepsilon L^2 =2\alpha\pi^2(1+3m^2),\qquad hL^2=8\pi^2\alpha m.
\end{align}
\end{subequations}
Both solutions give $\cE_\sigma=\varepsilon L - hM=\frac{2\pi^2n_\phi^2\alpha }{L} (1-m^2)$. As we depart from $\Omega_\text{c}$, we get deviations from the helix, which can be described by
\begin{subequations}
\label{eq:deviate}
\begin{align}
  \theta(x) &= \theta_0 + \vartheta(x)\\
  \phi(x) &= Qx + \varphi(x)
\end{align}
\end{subequations}
with $Q=2\pi/L$. In the case of the first solution \eqref{eq:sol_one}, these deviations describe the canting of the helix. Since $h=0$, the energy density $\varepsilon - hz(x)+P^2/(2\rho L^2)=\text{const.}$. Small deviations about the second solution correspond to the appearance of a spatially localized energy density.

$\Omega_\text{c}$ represents a first order transition at which the gradient of the dispersion jumps. To evaluate the jump, we return to the constrained minimization formulation given by \eqref{eq:const_min}, and evaluate the energy and solid angle to quadratic order in the deviations \eqref{eq:deviate}. The first order terms vanish since $\vartheta(x)$ has zero average, leaving the quadratic forms
\begin{multline}
  \cE_2 = \frac{\alpha}{2}\int dx\,\left[\vartheta'^2+Q^2\cos2\theta_0\vartheta^2\right.\\ \left.+\sin^2\theta_0\varphi'^2+2Q\sin2\theta_0 \varphi'\vartheta\right]
\end{multline}
\begin{equation}
\Omega_2 = \int dx\left[\frac{Q}{2}\cos\theta_0\vartheta^2+\sin\theta_0\vartheta\varphi'\right].
\end{equation}
Minimizing $\cE_2$ subject to fixed $\Omega_2$ involves the quadratic form $\cE_2-\mu\Omega_2$. The stationary points of this functional give a generalized eigenvalue problem for the Lagrange multiplier $\mu$. Writing
\begin{align}
  \vartheta(x) = \vartheta_0 e^{iQx}+\text{c.c}\\
  \varphi(x) = \varphi_0 e^{iQx}+\text{c.c}
\end{align}
gives a matrix eigenvalue problem, with eigenvalues $\mu = Q\cos\theta_0,\, 2Q\cos\theta_0$. Since $\mu=\partial\cE/\partial \Omega$ we arrive at the simple statement that the gradient in the dispersion jumps by a factor of $2$ at the transition to a localized solution (see Figure \ref{fig:Metastable}).

\subsection{Soliton Limit}\label{sec:soliton}

\begin{figure}
  \includegraphics[width = \columnwidth]{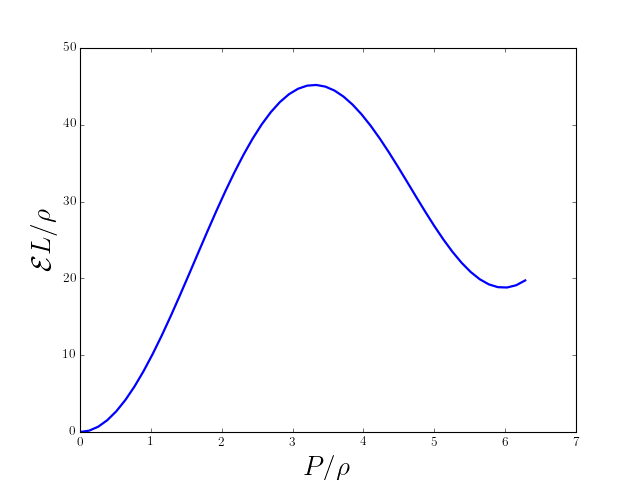}
  \caption{Soliton dispersion \eqref{eq:tw_dispersion} for $s=1/2$ and $\Delta=0.1$.}
  \label{fig:soliton}
\end{figure}

The opposite limit corresponds to almost complete polarization. In this situation, the magnetization axis will approach the $z$-axis as we move away from a localized configuration, so that we can regard the domain as infinite with the boundary conditions
\begin{equation}
  \mathbf{n}(\pm \infty) = \hat{\mathbf{z}}.
\end{equation}
To obtain this type of solution, the conserved quantities $p_\phi$, $p_\psi$, and $\varepsilon$ must be chosen so that the two largest roots $z_{1,2}$ of the effective potential of $V_\text{eff}(z)$ are at $z=1$, with the smallest root $z_3$ somewhere in $[-1,1]$. In this case $z=1$ is a point of unstable equilibrium, and an excursion to $z_3$ can occur at any $x$. The effective potential is then
\begin{equation}
  V_\text{eff}(z)=-h(z-z_3)(1-z)^2,
  \label{eq:Veff_soliton}
\end{equation}
giving rise to a soliton centered at $x=x_0$
\begin{equation}
	z(x) = z_{3}+\left(1-z_{3}\right)\tanh^{2}\left(\sqrt{\frac{h\left(1-z_{3}\right)}{2\alpha}}\left(x-x_{0}\right)\right).
  \label{eq:soliton}
\end{equation}
Using this solution, one can compute the angle $\phi(x)$, solid angle $\Omega$
\begin{equation}
  \Omega = \int_{0}^{2\pi} (1-\cos\theta)d\phi = 4 \arctan\sqrt{\frac{1-x_3}{1+x_3} },
\end{equation}
and the deviation from complete magnetization
\begin{equation}
  \Delta L \equiv \int_{-\infty}^\infty \left[1-\cos\theta\right]dx=2\sqrt{\frac{2\alpha(1-x_{3})}{h}}.
\end{equation}
Noting that for the potential \eqref{eq:Veff_soliton} $\varepsilon=h$, we can find the energy of this configuration
\begin{equation}
	\begin{split}
	\cE&=\varepsilon L -hM+ \frac{\rho P^2}{2\rho L}\\
	&=\frac{8\alpha}{\Delta L}\left[1-\cos \left(\frac{P}{2\rho s}\right)\right]  + \frac{ P^2}{2\rho L}
	\end{split}
  \label{eq:tw_dispersion}
\end{equation}
where $P=s\Omega$ on the principal branch of the dispersion. This dispersion relation is shown in Figure \ref{fig:soliton}

The solution \eqref{eq:soliton} appears in Ref.~\onlinecite{Tjon:1977} on the continuum limit of the Heisenberg chain, as well as repeatedly in the literature on elasticity (see e.g. Ref.~\onlinecite{Audoly:2010}). For our purposes, the dispersion relation \eqref{eq:tw_dispersion} illustrates a key point about the metastability of currents in a ferromagnetic condensate. The scale of the soliton is $\Delta$, meaning that as we approach full polarization ($\Delta\to 0$) the energy diverges like $\Delta^{-1}$. This illustrates that the energy barrier diminishes with decreasing polarization.

As $\Omega\to 0\mod 4\pi$, $x_3\to 1$ and the size of the soliton diverges, invalidating our assumption of an infinite domain. Instead, we expect the configurations of a finite system to connect to the helical configurations discussed in the previous sections at the critical solid angle $\Omega_\text{c}$ given by \eqref{eq:Omega_helix}. We next see how this happens by using the exact solution to treat the boundary conditions more carefully.

\subsection{Solution in terms of elliptic functions}

As discussed in Section \ref{sec:2_analogies}, the general solution for $z(x)$ is an elliptic function, obtained as an inverse of the elliptic integral~\cite{Whittaker:1970}.
\begin{equation}
  x = \int^z \frac{dz'}{\sqrt{P(z')}},
  \label{eq:elliptic_raw}
\end{equation}
where $P(z)=-2V_\text{eff}(z)/\alpha$ is a cubic polynomial. To put this elliptic function in standard form, we write
\begin{equation}
  z = 2\tilde z +c,
\end{equation}
with $c$ chosen so that the three roots of the cubic sum to zero
\begin{equation}
	\label{eq:cdef}
	c\equiv \frac{p_{\psi}^{2}/(2\alpha)+\varepsilon}{3h}.
\end{equation}%
Then \eqref{eq:elliptic_raw} becomes
\begin{equation}
	\label{SpinorCurrents_zIndef}
	x = \sqrt{\frac{\alpha}{h}}\int^{\tilde z} \frac{d\tilde z'}{\sqrt{4\left(\tilde z'-e_{1}\right)\left(\tilde z'-e_{2}\right)\left(\tilde z'-e_{3}\right)}}
\end{equation}
with real roots $e_{1}>e_{2}>e_{3}$ obeying $e_{1}+e_{2}+e_{3}=0$. As a result, $\tilde z(x)$ is
\begin{equation}
	\label{SpinorCurrents_zSol}
	\tilde z(x) = \wp\left(\sqrt{\frac{h}{\alpha}}x+\omega_{3}\right)
\end{equation}
where $\wp(x)$ is the Weierstrass elliptic function and $\omega_{3}$ is the (imaginary) half period associated with $e_{3}$
\begin{equation}
	\label{SpinorCurrents_HalfPeriod}
	\omega_{3}=-i\int_{-\infty}^{e_{3}}\frac{d\tilde z}{\sqrt{4\tilde z^{3}-g_{2}\tilde z-g_{3}}}.
\end{equation}
From the form of the effective potential \eqref{eq:Veff} we can identify the Weierstrass invariants
\begin{equation}
	\label{SpinorCurrents_WeierstrassInvariants}
	\begin{split}	g_{2}&=\frac{\left(p_{\psi}^{2}/2\alpha+\varepsilon\right)^{2}}{3h^{2}}+1-\frac{p_{\varphi}p_{\psi}}{h\alpha}
		\\
	g_{3} &= \frac{\left(p_{\psi}^{2}/2\alpha+\varepsilon\right)^{3}}{27 h^{3}}-\frac{1}{6}\frac{\left(p_{\psi}^{2}/2\alpha+\varepsilon\right)}{h}\left(\frac{p_{\varphi}p_{\psi}}{h\alpha}-1\right)\\
  &\qquad-\frac{1}{2}\frac{\left(\varepsilon- p_{\varphi}^{2}/2\alpha\right)}{h}.
	\end{split}
\end{equation}
The shift \eqref{eq:cdef} can be found from the magnetization using
\begin{equation}
	\label{SpinorCurrents_Magnetization}
	\begin{split}
		M = \int_{0}^{L} z\, dx &= 2\int_{0}^{L}dx\,\wp\left(\sqrt{\frac{h}{\alpha}} x+\omega_{3}\right) + cL\\
		&=-4 \sqrt{\frac{\alpha}{h}}  n_{T}\zeta(\omega_1)+cL\\
		&=L\left(c-\frac{2\zeta(\omega_1)}{\omega_{1}}\right)
	\end{split}
\end{equation}
where $\zeta(z)$ is the Weierstrass zeta function, defined as
\begin{equation}
  \zeta'(z)=-\wp(z)\qquad \zeta(z)\to \frac{1}{z}.
\end{equation}
Furthermore, if we have one period of the $\theta(x)$ motion in $L$ then
\begin{equation}
	\label{eq:PeriodExplicit}
	L = 2\sqrt{\frac{\alpha}{h}}\omega_{1}.
\end{equation}
We can then solve the $\phi(x)$ quantization condition in \eqref{eq:pbc} in terms of $h$ and $c$, and use the result to find the momentum and energy. Ref.~\onlinecite{Whittaker:1970} gives expressions for the angles $\phi(x)$, $\psi(x)$, from which we find
\begin{equation}
  i\left[\phi(L)-\phi(0)\right] = 2\left(\zeta(k)-\zeta(l)\right)\omega_{1}-2\zeta(\omega_1)(k-l)\label{eq:phi_sol}
\end{equation}
where $k$ and $l$ are the imaginary values of $x+\omega_{3}$ where $\theta=0$ and $\pi$. That is,
\begin{equation}
	\label{SpinorCurrents_kldef}
	\begin{split}
	\wp(l)&=\frac{1}{2}\left[1-c\right]\\
	\wp(k)&=-\frac{1}{2}\left[1+c\right].
	\end{split}
\end{equation}
Using \eqref{eq:phi_sol} in the quantization condition \eqref{eq:pbc} defines a curve in the $g_2-g_3$ plane that implicitly defines the dispersion relation $\cE(P)$.

\begin{figure}
  \includegraphics[width = \columnwidth]{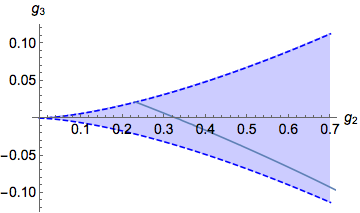}
  \caption{(Solid line) Values of the Weierstrass invariants $g_2$, $g_3$ for which the quantization conditions \eqref{eq:pbc} are satisfied for $m=0.3$. (Dashed lines) Curves of vanishing modular discriminant \eqref{eq:mod}: the top curve corresponds to the helix solution; the lower to the soliton. For larger values of $m$ the solid curve would lie very close to the soliton line, though it always starts on the helix line.}
  \label{fig:g2g3}
\end{figure}

As we have seen, the spin helix and the soliton correspond to the merging of the two lower and two upper roots respectively. This occurs when the \emph{modular discriminant} vanishes
\begin{equation}
  g_2^3-27g_3^2=0,
  \label{eq:mod}
\end{equation}
with $g_3=\pm\sqrt{27 g_2^3}$ corresponding to the helix and soliton respectively. In Figure \ref{fig:g2g3} we show the curve in this region of the $g_2-g_3$ plane that gives the correctly quantized solutions for $m=0.3$. The curve tracks the soliton branch quite closely despite the modest values of the polarization, but terminates on the helix branch. For larger values of the polarization the curve is essentially indistinguishable from the soliton branch. This is because the tail of the soliton is exponential, so at higher polarization we rapidly enter a regime where the quantization condition is satisfied to a very good approximation by the soliton.

Unfortunately, this also means that discriminating between the soliton and `true' finite $L$ solution becomes numerically difficult.
In the next section we will show that a direct numerical solution of the constrained minimization problem \eqref{eq:const_min} is an easier route to finding the dispersion relation. However, the above analysis does indicate a key difference between the finite size system and the soliton limit studied in the previous section. In the finite system the dispersion starts from a spatially extended helical solution at low $\Omega$ before a localized soliton-like configuration appears at $\Omega_\text{c}$ given by \eqref{eq:Omega_helix}.

\section{Numerical Calculation} \label{sec:numerical}

\begin{figure}
  \includegraphics[width = \columnwidth]{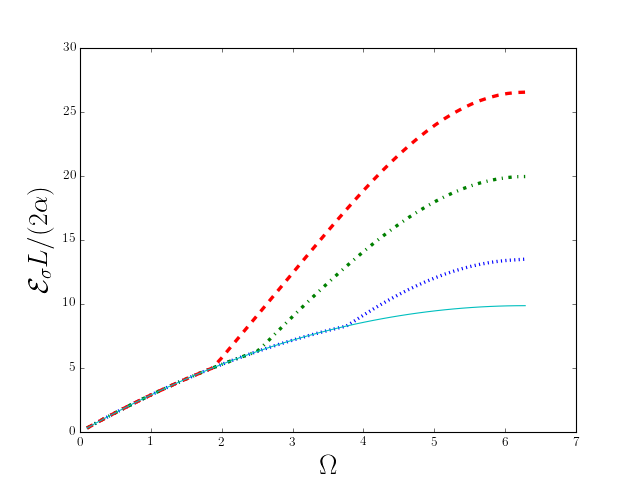}
  \caption{Numerically computed $\cE_\sigma$ for $m=0.4$ (dotted line), $0.6$ (dot dashed), and $0.7$ (dashed). Solid line shows the helix dispersion relation $ \cE_\sigma(\Omega) = \frac{2\pi\alpha\Omega}{L}\left(1-\frac{\Omega}{4\pi}\right)$.}
  \label{fig:Esigma}
\end{figure}

In Section \ref{sec:conserved} we saw that the calculation of the dispersion relation could be phrased in terms of the constrained optimization problem
\begin{equation}
  \cE_\sigma(\Omega, M) = \min_{\substack{\mathbf{n}(x)\\ \Omega[\mathbf{n}]=\Omega \\ M[\mathbf{n}]=M}} \frac{\alpha}{2}\int_0^L\left(\frac{d\mathbf{n}}{dx}\right)^2 dx.
  \label{eq:const_min2}
\end{equation}
In this section we perform the numerical calculation of $\cE_\sigma(\Omega, M)$ to verify the features that we have found analytically. To this end, it is convenient to parameterize the configurations using the stereographic variable
\begin{equation}
  \eta(x) = \tan(\theta(x)/2) e^{i\phi(x)}.
\end{equation}
In terms of $\eta(x)$, \eqref{eq:const_min2} becomes
\begin{align}
  \cE_\sigma(\Omega, M) &= \min_{\substack{\eta(x)\\ \Omega[\eta]=\Omega \\ M[\eta]=M}} 2\alpha \int_0^L dx\, \frac{|\eta'|^2}{(1+|\eta|^2)^2}\\
  M[\eta] &= \int_0^L dx\,\frac{1-|\eta|^2}{1+|\eta|^2}\\
  \Omega[\eta] &= i\int_0^L dx\,\frac{\eta\bar \eta'-\bar \eta \eta'}{1+|\eta|^2}.
  \label{eq:const_min_eta}
\end{align}
It is then straightforward to discretize each of these functionals. For the solid angle, we use the discretization
\begin{equation}
  \Omega = -i\sum_j \log\frac{1+\bar \eta_j \eta_{j+1}}{1+\eta_j \bar \eta_{j+1}},
  \label{eq:discrete_omega}
\end{equation}
which is based on the overlap of two spin-1/2 spinors
\begin{align}
  \Psi_j &= \frac{1}{\sqrt{1+|\eta_j|^2}} \begin{pmatrix}
  1\\ \eta_j
  \end{pmatrix}\\
  \braket{\Psi_j|\Psi_k}&= \frac{1+\bar\eta_j\eta_k}{\sqrt{(1+|\eta_j|^2)(1+|\eta_k|^2)}}.
\end{align}
\eqref{eq:discrete_omega} has the advantage of each term never being larger than $\pi$.

\begin{figure}
  \includegraphics[width = \columnwidth]{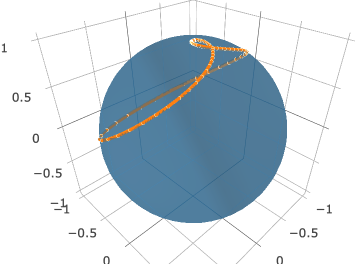}
  \caption{Example of a soliton-like configuration of $\mathbf{n}(x)$ for $m=0.9$, $\Omega =1.1\pi$.}
  \label{fig:sphere}
\end{figure}

Constrained optimization is performed using the SLSQP method, as implemented in \texttt{SciPy}~\cite{Jones:2016}. Results are shown in Figure \ref{fig:Esigma} for several values of the magnetization. As discussed in Section \ref{sec:disp}, the full dispersion relation in a finite system consists of two distinct regions. For small $\Omega \mod 4\pi$ we have a spin helix, and the dispersion follows the result \eqref{eq:helix_disp}. At $\Omega_\text{c}= 2\pi (1-\cos\theta)=2\pi (1-m)$ the helix  encircles the $z$-axis at constant latitude. Thereafter, the only way to increase $\Omega$ at fixed $M$ is by forming a localized, soliton-like solution (see Figure \ref{fig:sphere}), with the gradient of the dispersion jumping by a factor of two. As the polarization approaches unity, the dispersion tends to that of the soliton in an infinite system, with a diverging energy barrier.

\section{Discussion} \label{sec:conclude}

\begin{figure}
  \includegraphics[width = \columnwidth]{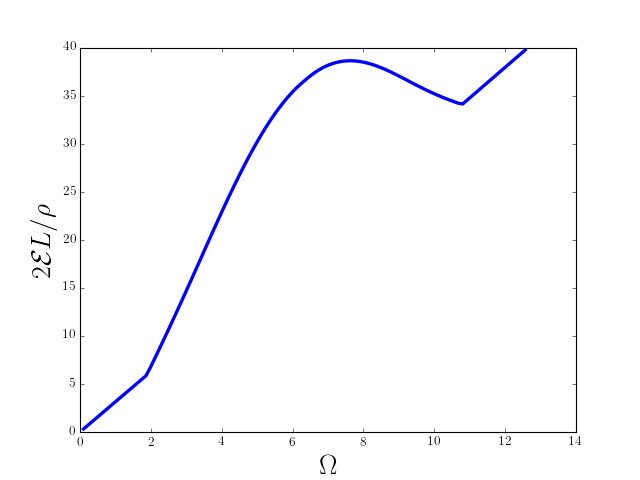}
  \caption{Full dispersion showing metastable minimum for the spin-1/2 case with $m=0.7$. The end point of the curve corresponds to $\cE=2\pi^2 \rho/L$, and corresponds to the energy a fully polarized state with a single flux quantum.}
  \label{fig:Metastable}
\end{figure}

In Section \ref{sec:conserved} we saw that the total energy has the form
\begin{equation}
\cE_p(\Omega,M) =  \frac{\rho}{2L} (2\pi p + s\Omega)^2 + \cE_\sigma(\Omega, M)
\label{eq:dispersion_total2}
\end{equation}
For concreteness, we now confine ourselves to the spin-1/2 case, where $\alpha=\rho/4$. In this case there is only one branch of the dispersion, and the helix part of the dispersion relation is piecewise linear
\begin{align}
\cE_p=\frac{\rho}{L}\left[2\pi^2 p^2 + (2p+1)\frac{\pi \Omega}{2}\right],\qquad 0\leq \Omega\leq 4\pi
\end{align}
In Figure \ref{fig:Metastable} we show the full form of the dispersion for $m=0.7$. The localized configurations give rise to energy barriers that are responsible for persistent currents.

A complete analysis of metastability would involve a treatment of the dynamics of a condensate that starts in a current carrying state. This is beyond the scope of this paper, but we offer some comments as to the form such a treatment must take. For the current to decay it must be possible for the condensate to lose momentum. There are two ways this may happen. The explicit translational invariance of the model may be broken e.g. by a nonuniform potential. Alternatively, the condensate may interact with excitations that are in thermal equilibrium in the lab frame, e.g. the normal component of a finite temperature condensate.

Since $P/(s\rho)=\Omega\mod 4\pi$ in our model, we then expect $\Omega$ to acquire some dynamics that may lead to the system passing over the maximum of the dispersion curve. Note that since this curve is the result of a constrained minimization problem, the maximum is a saddle point of the energy with one negative eigenvalue to the Hessian.

In the context of superconductors the classic Langer--Ambegaokar--McCumber--Halperin theory \cite{Langer:1967aa,McCumber:1970aa} provides a description of the decay of supercurrents in a scalar condensate. This takes the form of a time-dependent Ginzburg--Landau theory with a phenomenological noise term to drive the dynamics of the order parameter. The decay of persistent currents is due to the noise driven motion over the saddle point (or `phase slip') connecting two metastable minima. In the case of ultracold gases, a description based on the time-dependent Gross--Pitaevskii equation is more appropriate: see Ref.~\onlinecite{Mathey:2014aa} for recent work in this direction.

In a spinor condensate the magnetization provides a simple way to  vary the height of the energy barrier that must be overcome. In Ref.~\onlinecite{Beattie:2013aa} persistent currents were observed in a two-component Bose condensate confined to a toroidal trap. The gas was initialized in a fully polarized state with three quanta of circulation, before an RF pulse tips the magnetization $\mathbf{M}$ away from the $z$-axis. The key finding was a critical $z$-axis polarization of $m_\text{c}=0.64$, with the lifetime of current carrying state changing from $90\,\text{s}$ to below $20\,\text{s}$ as the polarization is reduced.

From Figure \ref{fig:Metastable} we see that even at $m=0.7$, although there is still a metastable minimum in the dispersion, the maximum lies below the energy $\cE=2\pi^2 \rho/L$ of a single flux quantum in a fully polarized state. There is therefore enough energy in such an initial state to pass over the energy barrier. Given the simplifications of our model (1D, incompressible condensate) and the absence of any treatment of dynamics, this disagreement is perhaps not too surprising. The advantage of our approach is that it provides a simple model of persistent currents that applies even in the incompressible limit where a scalar condensate would flow forever.


The author thanks Stefan Baur and Nigel Cooper for discussions at an early stage of this work.


%

\end{document}